\documentclass[reprint,aps,prl,superscriptaddress,amsmath,amssymb]{revtex4-1}
\usepackage{graphicx}% Include figure files
\usepackage{bm}% bold math
\usepackage{braket} %Dirac notation
\usepackage{bbold}
\usepackage[colorlinks=true,urlcolor=blue,citecolor=blue,linkcolor=blue,bookmarks=false,pdfstartview={FitH}]{hyperref}
\newcommand{\be}{\begin{equation}}
	\newcommand{\ee}{\end{equation}}

\newcommand{\bea}{\begin{eqnarray}}
	\newcommand{\eea}{\end{eqnarray}}

\newcommand{\e}{\varepsilon}

\renewcommand{\vec}[1]{{\bf #1}}
\def\nn{\nonumber\\}

\begin{document}
	\title{Chiral Spin Mode on the Surface of a Topological Insulator}

	\author{H.-H.~Kung}
	\email{skung@physics.rutgers.edu}
	\affiliation{Department of Physics \& Astronomy, Rutgers University, Piscataway, New Jersey 08854, USA}
	\author{S.~Maiti}
	\affiliation{Department of Physics, University of Florida, Gainesville, Florida 32611, USA}
	\author{X.~Wang}
	\affiliation{Department of Physics \& Astronomy, Rutgers University, Piscataway, New Jersey 08854, USA}
	\affiliation{Rutgers Center for Emergent Materials, Rutgers University, Piscataway, New Jersey 08854, USA}
	\author{S.-W.~Cheong}
	\affiliation{Department of Physics \& Astronomy, Rutgers University, Piscataway, New Jersey 08854, USA}
	\affiliation{Rutgers Center for Emergent Materials, Rutgers University, Piscataway, New Jersey 08854, USA}
	\author{D. L.~Maslov}
	\email{maslov@phys.ufl.edu}
	\affiliation{Department of Physics, University of Florida, Gainesville, Florida 32611, USA}
	\affiliation{National High Magnetic Field Laboratory, Tallahassee, Florida 32310, USA}
	\author{G.~Blumberg}
	\email{girsh@physics.rutgers.edu}
	\affiliation{Department of Physics \& Astronomy, Rutgers University, Piscataway, New Jersey 08854, USA}
	\affiliation{National Institute of Chemical Physics and Biophysics, 12618 Tallinn, Estonia}

	\begin{abstract}
		Using polarization-resolved resonant Raman spectroscopy, we explore collective spin excitations of the chiral surface states in a three dimensional topological insulator, Bi$_2$Se$_3$.
		We observe a sharp peak at 150\,meV in the pseudovector $A_2$ symmetry channel of the Raman spectra.
		By comparing the data with calculations, we identify this peak as the transverse collective spin mode of surface Dirac fermions.
		This mode, unlike a Dirac plasmon or a surface plasmon in the charge sector of excitations, is analogous to a spin wave in a partially polarized Fermi liquid, with spin-orbit coupling playing the role of an effective magnetic field.
	\end{abstract}
	\maketitle
	%
	%\subsection{Introduction}
	Magnets and partially spin-polarized Fermi liquids support collective spin excitations (spin waves), in which all electron spins respond coherently to external fields, and the ``glue'' that locks the phases of precessing spins is provided by the exchange interaction.
	In nonmagnetic materials where inversion symmetry is broken but time-reversal invariance remains intact, strong spin-orbit coupling (SOC) may play the role of an effective magnetic field, which locks electron spins and momenta into textures.
	This phenomenon is encountered in three-dimensional (3D) topological insulators (TIs), which harbor topologically protected surface states \cite{Fu2007A,Zhang2009,Chen2009,Hsieh2008,Hasan2011}.
    These states have been a focus of intense studies, both from the fundamental point-of-view
	\cite{Bansil2016,Wang2013ScienceGedik,Zhu2014,Wu2016Science,Jozwiak2016,Li2016SciRep,Shao2017NanoL} 
	and for potential applications in spintronics devices
\cite{Pesin2012NMater,Xu2014NPhys,Fan2016NNano,Fan2016NNano,Wang2015PRL,Scharf2016PRL,Kondou2016NPhys,Tian2017SciAdv,Ando2017JPSJ}.
	However, the many-body interactions leading to collective effects in TIs remain largely unexplored.
	An essential aspect of this physics is an interplay between the Coulomb interaction and SOC, which is expected to give rise to a new type of collective spin excitations -- chiral spin waves \cite{Shekhter2005,Ashrafi2012,Maiti2015,Maiti2016,Maiti2017,Kumar2017}.
    In the long wavelength ($q=0$) limit, these modes are completely decoupled from the charge channel and thus distinct from spin-plasmons \cite{Raghu2010,Kondo2013}, Dirac plasmons \cite{Pietro2013,Palitano2015}, and surface plasmons in TIs \cite{Kogar2015,Palitano2015}. 
    
\begin{figure*}
		\includegraphics[width=11.5cm]{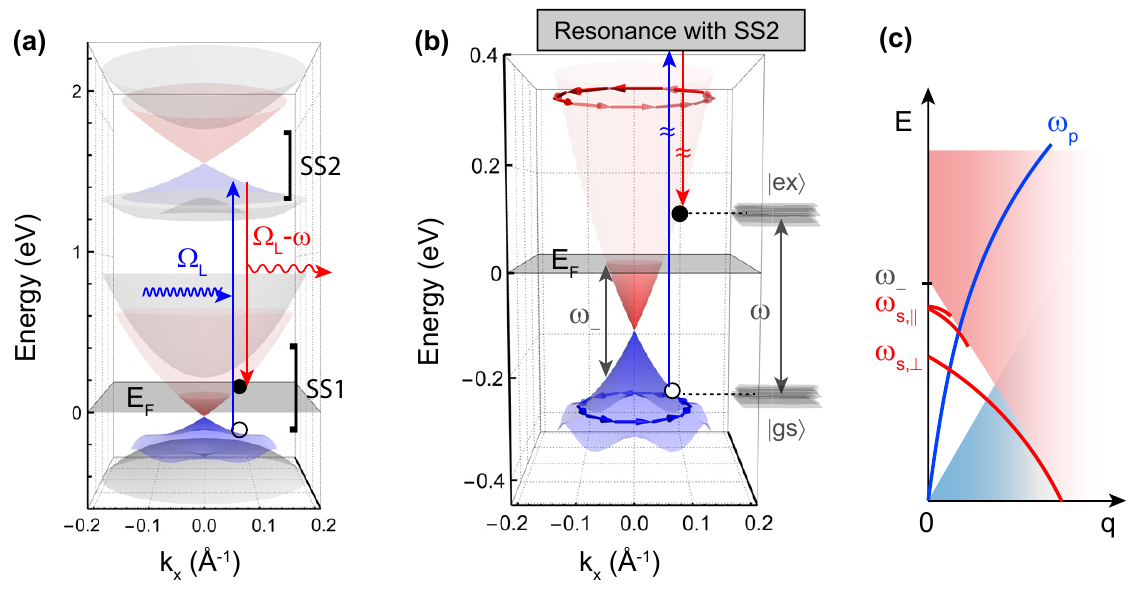}
		\caption{\label{Fig1} 
			%\textbf{The band structure and collective modes in Bi$_2$Se$_3$.}
			(a)~The band structure of Bi$_2$Se$_3$ around the Brillouin zone center, 
			reconstructed from ARPES results in Refs.\,\cite{Nomura2014,Sobota2013}.
			The low-energy Dirac cones of surface states (SS1) are described by Eq.\,(\ref{eq:1}) 
			with parameters $m^\ast\approx 0.066\,\text{eV}^{-1}\text{\AA}^{-2}$, $v_1\approx 2.4$\,eV\AA, and $v_{\text{w}}\approx 25$\,eV\AA$^3$ \cite{SM}.
			The upper and lower Dirac cones of opposite chirality are shown in red and blue, whereas the bulk bands are shown in gray. 
			A pair of unoccupied surface Dirac cones (SS2)
			resides about 1.8\,eV above SS1 \cite{Sobota2013}.
			The arrows illustrate a resonant Raman process stimulated by an incoming photon	with energy $\Omega_L$, in which an electron is promoted from the lower to the upper Dirac cone of SS1, through resonant scattering via the intermediate states, SS2.
			The energy difference between the excited and ground state defines the Raman shift, $\omega$. 
			(b)~Enlarged view of SS1 around $E_F$ with arrows showing the spin textures, where $\omega_-$ is the threshold energy for direct transitions between occupied states in the lower Dirac cone and available states of the upper cone.
			(c)~Schematic dispersions of the particle-hole continua	in the spin (red) and charge (blue) channels, Dirac plasmon \cite{Raghu2010,Pietro2013,Palitano2015} (blue line), and 
			chiral spin modes \cite{Maiti2015} (red lines).  
			The $\omega_{s,\perp}$ mode is observed in this work.
		}
	\end{figure*}

	In this Letter, we employ polarization-resolved resonant Raman spectroscopy, a technique of choice for probing the collective charge \cite{Cardona1982,Devereaux2007}, spin \cite{Cottam1986,Gozar2004,Perez2016,Gretarsson2016PRL} and orbital excitations \cite{Miyasaka2005PRL}, to study collective spin excitations of the chiral surface states in Bi$_2$Se$_3$.
	To enhance the signal from the surface states, we tune the energy of incoming photons in resonance with a transition between two sets of chiral surface states: near the Fermi energy and about $1.8$\,eV above it [Fig.~\ref{Fig1}(a)] \cite{Sobota2013}. 
	We observe a long-lived excitation at 150\,meV in the 
	pseudovector symmetry channel of the Raman spectra, which is most pronounced at low temperatures but persists up to room temperature. 
	By comparing the data with calculations, we identify this excitation as the transverse collective chiral spin mode supported by spin-polarized surface Dirac fermions. 
	Such collective modes -- first predicted for non-topological systems \cite{Shekhter2005,Ashrafi2012,Maiti2015,Maiti2016,Perez2016,Maiti2017,Kumar2017} but hitherto unobserved -- are ``peeled off'' from the continuum of particle-hole excitations by the exchange interaction.
	
	Chiral surface states in a 3D TI are described by the Hamiltonian \cite{Fu2009}:
	\begin{equation}\label{eq:1}
		\hat H({\bf k}) = \frac{k^2}{2m^\ast} \hat\sigma_0+v_1 \boldsymbol{\hat\sigma}\cdot\vec{\tilde k}\,,
	\end{equation}
	where $m^\ast$ is the effective mass,  
	$\boldsymbol{\hat\sigma}=(\hat\sigma_1,\hat\sigma_2,\hat\sigma_3)$ are the Pauli matrices, $\hat\sigma_0$ is the 2$\times$2 unit matrix, 
    and $\vec{\tilde k}=(k_y,-k_x,\frac{v_{\text{w}}}{v_1}[k_+^3+k_-^3])$ with $k_\pm\equiv k_x\pm i k_y$. 
	The $z$ component of $\vec{\tilde k}$ describes hexagonal warping of the surface states away from the Dirac point \cite{Fu2009}. 
	The spectrum of Eq.\,(\ref{eq:1}) consists of hexagonally warped electron- and hole-like Dirac cones of opposite chirality.
	A light-induced excitation from the occupied state in the hole cone to an empty state in the electron cone
	is accompanied by a spin-flip of the quasi-particle [Fig.~\ref{Fig1}(b)].
	Such direct transitions form a continuum which starts at the threshold energy $\omega_-$ [Fig.~\ref{Fig1}(c)] \cite{Riccardi2016,Kashuba2009}.
	
	Due to the Pauli exclusion principle, two electrons in the triplet state avoid each other, thus reducing the energy of the Coulomb repulsion. 
	Therefore, the repulsive Coulomb interaction between electrons translates into an attractive exchange interaction between their spins, leading to bound states below the continuum of spin-flip excitation, i.e., chiral spin modes.
	In general, there are three such modes [red curves in Fig.~\ref{Fig1}(c)], which correspond to linearly polarized oscillations of the magnetic moment in the absence of the external magnetic field \cite{Maiti2015,Maiti2016,Maiti2017}. 
	At $q=0$, there is a doubly degenerate mode with an in-plane magnetic moment and a transverse mode with an out-of-plane moment, with energies $\omega_{s,||}$ and $\omega_{s,\perp}$, respectively.
	Because the chiral-spin modes are below the continuum, they are expected to remain long-lived even at elevated temperatures \cite{Maiti2015PRL}. 
	These modes are in essence (zero-field) 
	spin waves that can be measured by resonant Raman scattering because they couple to the electromagnetic field through antisymmetric Raman tensors \cite{Maiti2017}.
	
	%\subsection{Experimental Results}
	\begin{figure}
		\includegraphics[width=7cm]{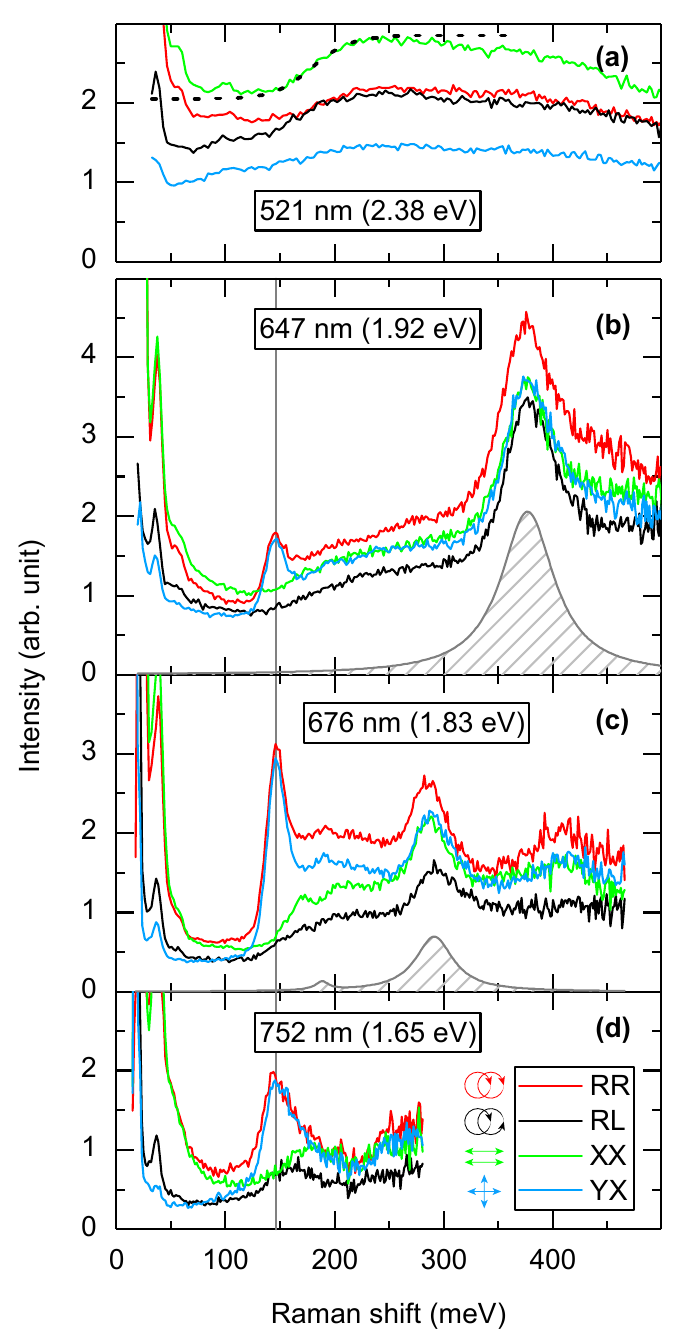}
		\caption{\label{Fig2} 
			%\textbf{Effect of resonance on the $150$\,meV peak in the Raman intensity.} 
			The intensities 
			of secondary emission
			are measured in circular and linear (with respect to crystallographic axes) scattering polarization geometries, as shown pictorially next to the figure legend in panel (d) and also defined in Supplementary \cite{SM},
			at 24\,K with (a)~521\,nm (2.38\,eV), (b)~647\,nm (1.92\,eV), (c)~676\,nm (1.83\,eV) and (d)~752\,nm (1.65\,eV) excitations. 
			The vertical solid gray line indicates the sharp 150\,meV peak, most pronounced for the incident photon energy of 1.83\,eV, which is the closest to the energy difference between SS1 and SS2 [Fig.~\ref{Fig1}(a)]. 
			The dotted black line in (a) is a guide to the eye showing the energy threshold for the surface-to-bulk excitations \cite{SM}.
			The hatched	areas in (b) and (c) are Lorentzian fits to the excitonic photoluminescence peaks.
		}
	\end{figure}
	
	Bi$_2$Se$_3$ is an archetypical 3D TI, with a rhombohedral crystal structure with $D_{3d}^5$ group symmetry ($R\overline{3}m$) in the bulk. 
	The crystal is composed of quintuple layers weakly bonded by van der Waals force, allowing easy cleavage of optically flat $ab$ surfaces with the symmorphic $P6mm$ wallpaper group symmetry (2D point group $C_{6v}$) \cite{Li2013Mar,Slager2013}.
	Bi$_2$Se$_3$ has a relatively simple band structure near the $\Gamma$-point [Fig.~\ref{Fig1}], with a pair of topologically protected surface Dirac cones, labeled SS1, and another pair at about 1.8\,eV above SS1, labeled by SS2 [Fig.~\ref{Fig1}(a)].
	As-grown Bi$_2$Se$_3$ is usually electron-doped due to naturally formed Se vacancies \cite{Huang2012Aug}.
	In this study, we use well-characterized samples with low concentration of impurities and crystalline defects \cite{Dai2016}. 
    All the bulk phonon modes in this crystal are sharp with no signatures of any impurity modes, and all the surface phonon modes 
   are clearly observed \cite{Kung2017}. 
     The Fermi energy ($E_F$) was determined by scanning tunneling spectroscopy to be about 150\,meV above the Dirac point of SS1 [Fig.\,\ref{Fig1}(b)] \cite{Dai2016}. 
	
	The polarized Raman spectra were acquired in a quasi-backscattering geometry from the $ab$ surface of Bi$_2$Se$_3$ single crystals grown by modified Bridgman method \cite{SM}.
	We use 521, 647, 676 and 752\,nm lines of a Kr$^+$ laser for excitation, where the spot size is roughly $50\times 50\,\mu m^2$, and the power is about 10\,mW. 
	The scattered light is analyzed by a custom triple-grating spectrometer.
	
	In Fig.~\ref{Fig2}, we show spectra 
	of secondary emission
	for four scattering geometries employing both linear and circular polarizations, as defined in the Supplementary \cite{SM}.
	Of the four excitations, the 521\,nm (2.38\,eV) one is the farthest from 
	near-resonant transition between SS1 and SS2,
	while the 676\,nm (1.83\,eV) one is the closest. 
	The spectra contain contributions from electronic Raman scattering and exciton photoluminescence.
	The latter is present for all polarizations and can be subtracted from the spectra \cite{SM}.
	The signal below 50\,meV is dominated by phonon modes which are discussed elsewhere \cite{Kung2017}.
	
	%\subsection{Discussion}
	\begin{figure}
		\includegraphics[width=7cm]{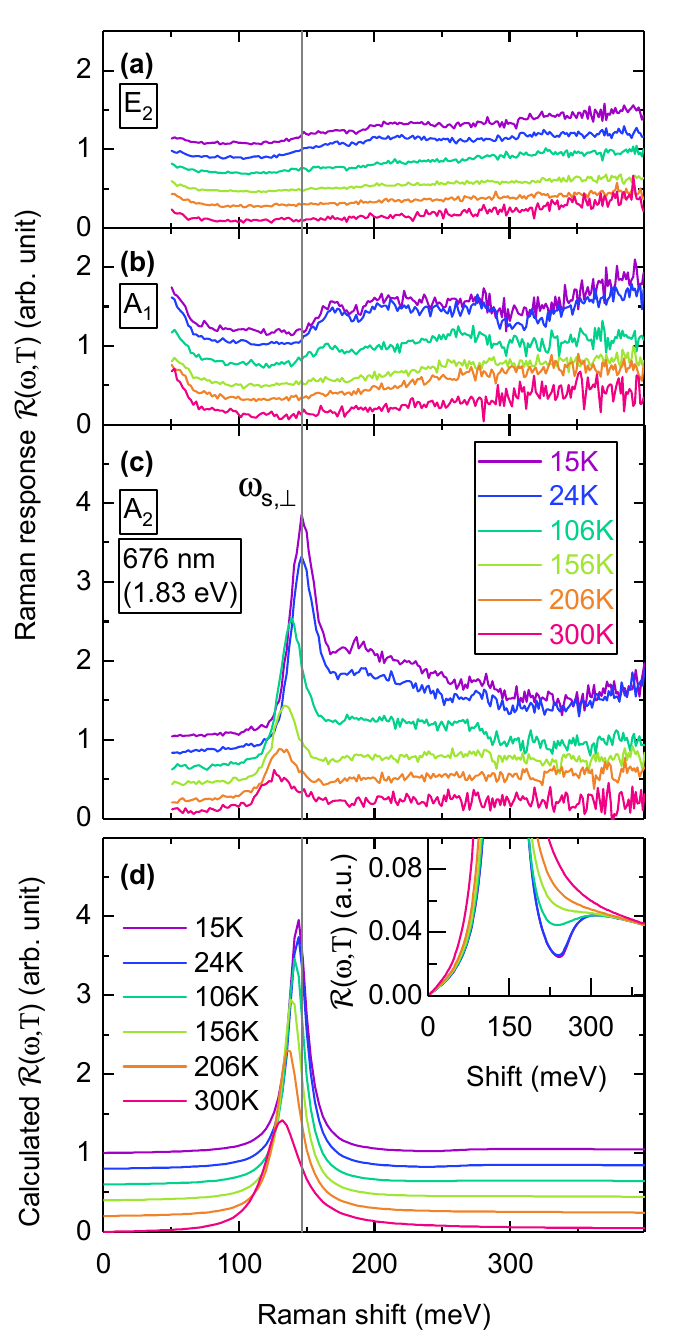}
		\caption{\label{Fig3} 
			Temperature dependence of the symmetry-separated Raman response $\mathcal{R}(\omega,T)$ in the 
			(a)~$E_2$, (b)~$A_1$ and (c)~$A_2$
			symmetry channels, measured with the near-resonant 676\,nm (1.83\,eV) excitation. 
			The photoluminescence background was subtracted from the raw data \cite{SM}. 
			The gray solid line locates the position of the sharp peak which is present only in the $A_2$ 
			channel.
			(d)~Calculated $\mathcal{R}(\omega,T)$ in the $A_2$ 
			channel
			with the dimensionless 
			interaction coupling constant
			$u\approx 0.6$
			and impurity broadening of $\Gamma= 8$\,meV was extracted form the data and 
			used 
			in the calculation.
			The lines are shifted vertically for clarity. 
			Inset: Zoom-in of the calculated $\mathcal{R}(\omega,T)$ without vertical shift, showing the spin-flip continuum with a threshold energy of 260\,meV.
		}
	\end{figure}
	
	The spectrum for the non-resonant 521\,nm (2.38\,eV) excitation shows no sharp peaks but a broad feature \cite{SM}. 
    This is in stark contrast to the spectra of other three excitations, where a sharp peak around 150\,meV is observed in the XY and RR geometries.
	The peak is the strongest for the 676\,nm (1.83\,eV) excitation, which is in near-resonance with the transition between 
	SS1 
	and
	SS2, 
	thus confirming the surface origin of the observed signal.
	In order to better understand the origin of the 150\,meV peak, we subtract the photoluminescence contributions and then utilize the symmetry properties of the Raman tensors to separate the measured spectra into the $E_2$, $A_1$ and $A_2$ symmetry channels of $C_{6v}$ point group \cite{SM}.
	In Fig.~\ref{Fig3}, we plot the temperature dependence of Raman response 
	$\mathcal{R}(\omega,T)$ 
	in 
	three symmetry channels.
	It is  clearly seen that the 150\,meV peak is associated with the $A_2$ symmetry channel. 
	The continuum broadens and becomes invisible above 150\,K, but the peak is still well-defined even at 
	$T=300$\, K.
	
	The basis functions of the $A_2$ representation of $C_{6v}$ transform as the $z$ component of the angular momentum, which is a pseudovector \cite{Ovander1960,Koster1963}.
	This suggests that the observed peak in the $A_2$ channel corresponds to a spin mode [marked by $\omega_{s,\perp}$ in Fig.~\ref{Fig1}(c)]
	with an out-of-plane magnetic moment (also a pseudovector).
	
	\begin{figure}
		\includegraphics[width=7cm]{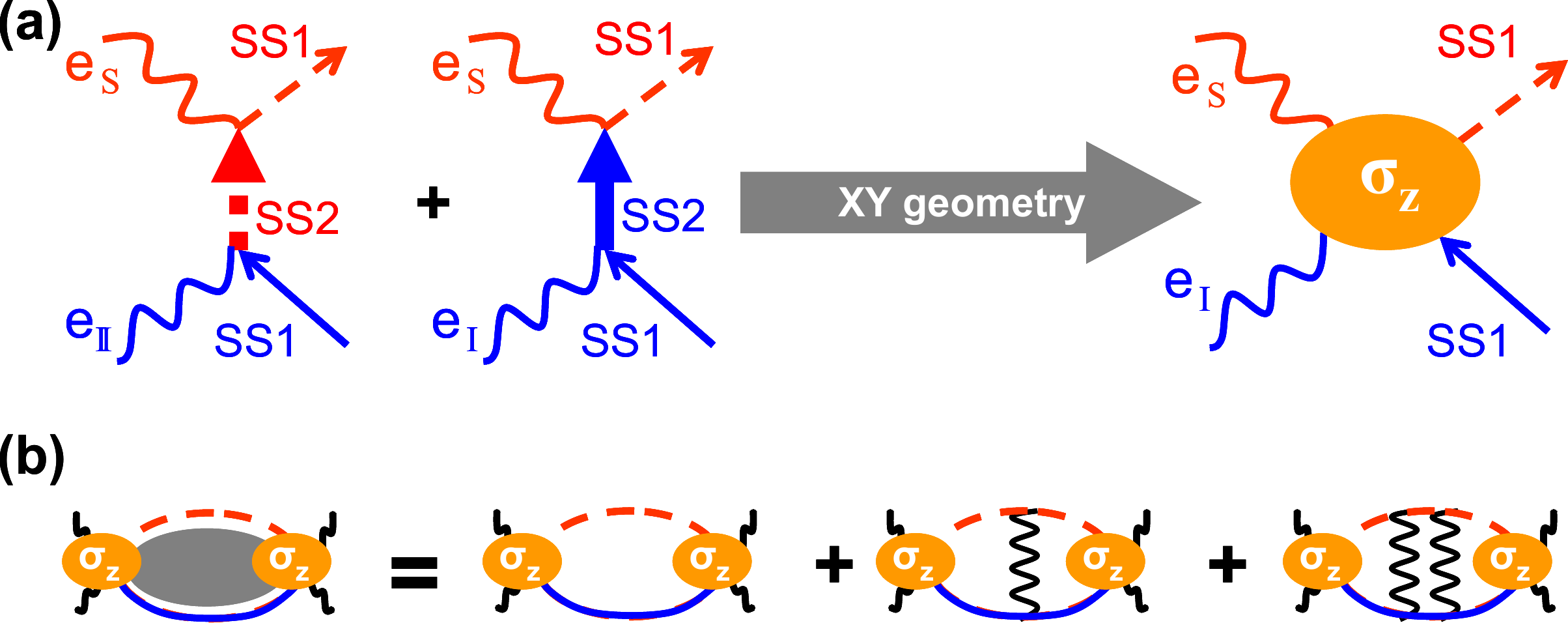}
		\caption{\label{FigVertex} 
			(a)~Raman vertex for resonant transitions from the lower cone in SS1 to the upper (left) and lower (right) cones in SS2.
			(b)~The 
			scattering cross-section
			dressed with vertex corrections to account for
			many-body effects.}
	\end{figure}
	
	To quantify the assignment of the 150\,meV peak to the out-of-plane chiral spin mode,  we calculate the Raman response of 
	surface chiral states.
	We are interested in
	spin-flip
	resonant Raman 
	processes between states near the Fermi level in SS1 and the states in SS2.
	Two resonance transitions are possible: 
	an electron from the lower cone of SS1 can be transferred into either the lower or upper cones of SS2 and come back to the upper cone of SS1, 
	producing a particle-hole pair. 
	The diagrams for the corresponding Raman vertex and scattering cross-section are shown in Figs.~\ref{FigVertex} (a) and (b), correspondingly.
	Since the shift in the photon energy $\approx 150$ meV is much smaller than energy difference between the Dirac points $E_g\approx 1.8$\,eV,
	the resonant part of the Raman vertex can be written as
	\bea\label{eq:DM1}
	\gamma
	({\bf k})\
	=\frac{\left({\bf e}_S\cdot{\bf p}_{u_1 u_2}\right)\left({\bf e}_I\cdot{\bf p}_{u_2l_1}\right)}{E_{u_2}({\bf k})
		-E_{l_1}({\bf k})-\Omega_L}+\frac{\left({\bf e}_S\cdot{\bf p}_{u_1 l_2}\right)\left({\bf e}_I\cdot{\bf p}_{l_2l_1}\right)}{E_{l_2}({\bf k})
		-E_{l_1}({\bf k})-\Omega_L},\nonumber\\
	\label{eq:gamma}
	\eea
	where $l_{1,2}$ and $u_{1,2}$ refer to the lower and upper cones of SS1 (SS2), 
	${\bf p}_{ab}$ is the matrix element of 
	a dipole transition between states $a$ and $b$, and ${\bf e}_{I,S}$ are the polarizations of incident and scattered photons, correspondingly.
	Furthermore, $E_{u_1,l_1}=\varepsilon^{(1)}_{\pm}({\bf k})$ and $E_{u_2,l_2}=E_g+\varepsilon^{(2)}_{\pm}({\bf k})$, where $\varepsilon^{(1)}_{\pm}({\bf k})=k^2/2m^{*}\pm v_1 \tilde k
	$ 
	are the eigenergies of Eq.~(\ref{eq:1}),  and $\varepsilon^{(2)}_{\pm}({\bf k})$ are the eigenenergies of a similar Hamiltonian for SS2.
	The scattering cross-section contains $|\gamma({\bf k})|^2$, integrated over $k\geq k_F$.
	However, if trigonal warping is neglected, the denominators of the first (second) terms in Eq.~(\ref{eq:gamma}) become $E_g+(v_{2}\pm v_1)k-\Omega_L$, where $v_{1,2}$ are the velocities of Dirac fermions in SS1 and SS2. 
	A characteristic feature of Bi$_2$Se$_3$ is that the lower cones
	of SS1 and SS2 are almost perfectly nested:  
	a fit to the ARPES data in Ref. \cite{Nomura2014,Sobota2013} gives $v_1=2.4$\,eV\AA\, and $v_2=2.0$\,eV\AA\,.
	Therefore, the second (hole-to-hole) term in $\gamma$
	is essentially dispersionless, and the corresponding transition probability is enhanced by a factor of $\approx 1/(E_g-\Omega_L)^2$, whereas the first (hole-to-electron) term is small~\footnote{If $\gamma$ disperses with $k$, the Raman response does not reduce to the spin susceptibility. However, the pole corresponding to the collective mode in the Raman response is still the same as in the spin susceptibility.}.
	This explains why only one resonance is observed in the experiment.
	
	Since 
	the
	initial and final states of the Raman vertex form a $2\times 2$ space,
	$\hat\gamma$ 
	can be expanded over a complete set of Pauli matrices as $\hat\gamma=c\hat\sigma_0+{\bf s}\cdot\boldsymbol{\hat\sigma}$. 
	In the $XY$ scattering geometry,
	which contains the $A_2$ symmetry channel of $C_{6v}$ group, ${\bf s}={\bf e}_{I}\times {\bf e}_{S}={\bf e}_z$ and thus $\hat\gamma\propto \hat\sigma_z$. 
	The Raman response function can then be written as\bea\label{eq:9}
	\mathcal{R}
	(\omega,T)\propto 
	\chi''_{zz}(\omega,T)/(E_g-\Omega_L)^2,\eea
	where
	$\chi''_
	{\alpha\beta}
	$ is the imaginary part of the
	$\alpha\beta^{\text{th}}$ 
	component of the spin susceptibility tensor. 
	The many-body interactions are accounted for
	within the Random Phase Approximation (RPA) with a Hubbard-like interaction ($U$) in
	the spin channel
	\cite{Maiti2017}:
	\be \label{eq:RPA}
	\hat\chi(\omega,T)=
	-\hat\Pi(\omega,T) \left(\mathbb{1}+\frac{U}{2}\hat\Pi(\omega,T) \right)^{-1} 
	\ee
	where 
	$\hat\Pi(\omega,T) $ 
	is
	obtained 
	by
	analytic
	continuation of 
	\be\label{eq:Pi}
	\Pi_{\alpha\beta}(i\omega_n)=T\sum_{\epsilon_m} \int_{\vec k}{\rm
		Tr}\left[\hat\sigma_\alpha\hat G
	_\vec k(i\epsilon_m+i\omega_n)
	\hat\sigma_\beta \hat G
	_\vec k
	(i\epsilon_m)\right].\nn
	\ee
	Here, $\int_\vec k\equiv \int\frac{d^2k}{(2\pi)^2}$, 
	$\hat G^{-1}
	_\vec k
	(i\epsilon_m)=
	i\e_m-\hat H(\vec k)+E_F+i\text{sgn}(\epsilon_m)\Gamma/2
	$,
	$\hat H(\vec k)$  
	is given by Eq.\,(\ref{eq:1}), and $\Gamma$ is impurity broadening \footnote{For calculation purposes, a momentum cutoff of $\Lambda_k=0.3$\,\AA$^{-1}$ was chosen \cite{SM}.}.
	$\chi_{zz}(\omega,T)$
	has a continuum of spin-flip excitations and a pole 
	which corresponds to the transverse collective mode.
	A simple result for the frequency of this mode can be obtained if one neglects hexagonal warping
	and considers
	the weak-coupling limit.
	In this case,
	$\omega_{s, \perp} = 
	2E_F \left[1-2 \exp(-4/u)\right]$,
	where $u\equiv UE_F/2\pi \hbar v_1^2\ll1$ is the dimensionless coupling constant.
	
	For a more general case,
	which includes 
	the realistic band structure 
	and finite temperature,
	the Raman response has to be evaluated numerically.
	The results of this calculation are shown in Fig.~\ref{Fig3}(d). 
	With the band structure parameters obtained from ARPES measurements \cite{Nomura2014}, the only
	two
	fitting parameters are the exchange coupling constant, fixed at $u\approx 0.6$ 
	to reproduce the mode 
	frequency
	at 15\,K,
	and the impurity scattering rate chosen as $\Gamma=8$\,meV.
	Comparison of the measured and computed spectra [Fig.~\ref{Fig3}(c) and (d), correspondingly] shows that the 
	model describes well not only the overall shape of the signal but also its evolution with temperature \cite{SM}.
    In particular, the theory reproduces the observed decrease in the peak position with increasing temperature, which can be ascribed to thermal smearing of the continuum boundary.
    For the well characterized samples studied here, the threshold of the spin-flip continuum is expected at $\omega_-\approx 260$\,meV \cite{SM}.
	However, the onset of this continuum 
	is difficult to observe because its spectral weight is 
	transferred into the collective mode. 
	In the inset of Fig.~\ref{Fig3}(d), we show a zoom into the 
	computed crossover region between the collective mode and the continuum. 
	A quantitative agreement between the theory and experiment gives us confidence in that the observed 150\,meV sharp peak in the $A_2$ symmetry channel is indeed a transverse chiral spin mode.
	
	%\subsection{Conclusion}
	In conclusion, 
	our results provide strong evidence for a new collective mode -- the transverse chiral spin wave -- in a time-reversal invariant system,
	a 3D TI.
	Strong spin-orbit coupling plays the role of a very high effective magnetic field, which protects the long-lived spin excitation.
	Such a robust collective spin mode may have
	potential applications in spintronics \cite{Najmaie2005,Dumas2014}, magnonics \cite{Kruglyak2010,Chumak2015,Urazhdin2014NNano}, optoelectronics \cite{Politano2017} and quantum computing \cite{Fu2008,DasSarma2008,Alicea2012}.
	Moreover, the present results demonstrate an efficient way of probing the dynamical response of Dirac fermions and their collective modes through optical measurement.
	The methods we use here pave a new route for discriminating bulk excitations from the surface modes and for exploring collective properties of chiral fermions.
	%%%%%%%%%%%%%%%%%%%End of main text%%%%%%%%%%%%%%%%%%%%%%%%%%%%
	
	\begin{acknowledgments}
		We are grateful to B. S.~Dennis and A.~Lee for technical support,
		and to T.~P.~Devereaux, A.~F.~Kemper, P.~Lemmens, 
		R.~Merlin and J.~A.~Sobota for stimulating discussions. 
		G.B. and H.-H.K. acknowledge support from NSF Grant No.~DMR-1709161. 
        S.-W.C. and X.W. acknowledge support from NSF Grant No.~DMREF-DMR-1629059. 
		D.L.M. acknowledges support from UF DSR  Opportunity Fund
		OR-DRPD-ROF2017.
		G.B. also acknowledges partial support from QuantEmX 
		grant from ICAM, the Gordon and Betty Moore 
		Foundation through Grant GBMF5305, and from the 
		European Regional Development Fund project TK134. 
	\end{acknowledgments}
	
	%\bibliography{Kung2017}

\clearpage
\section*{Supplemental Material for:\\
	Chiral Spin Mode on the Surface of a Topological Insulator}
\renewcommand{\thefigure}{S\arabic{figure}}
\renewcommand{\theequation}{S\arabic{equation}}
%%%%%%%%%%%%%%%%%%%%%%
\subsection{I.~Material \& Methods} \label{sec:supp.i}
\subsubsection{Material preparation}
The single crystals measured in this spectroscopic study were grown by modified Bridgman method.
Mixtures of high-purity bismuth (99.999\%) and selenium (99.999\%) with the mole ratio $\rm Bi:Se=2:3$ were heated up to 870\,$^\circ$C in sealed vacuum quartz tubes for 10 hours, and then slowly cooled to 200\,$^\circ$C with rate 3\,$^\circ$C/h, followed by furnace cooling to room temperature.

The Bi$_2$Se$_3$ crystals used in this study were characterized by STM and phononic Raman scattering studies in Refs.\,\cite{Dai2016,Kung2017}. 
The Fermi energy ($E_F$) is determined by scanning tunneling spectroscopy to be about 150\,meV above the Dirac point of SS1 [Fig.\,1(b) in Main Text] \cite{Dai2016}.
Characterization confirmed that the samples have low concentration of impurities, Se vacancies, or other crystalline defects. 
All the bulk phonon modes in this crystal are sharp with no signatures of impurity modes, and all the expected surface phonon modes are clearly observed \cite{Kung2017}. 
All spectroscopic features we present in this study were reproducible for a series of cleaves, immediately observed for each cool down, and did not show any signatures of time-dependent contamination.

\subsubsection{Raman scattering}
In this study, we used the 520.8, 647.1, 676.4 and 752.5\,nm lines of a Kr$^+$ ion laser to promote secondary emission from the Bi$_2$Se$_3$ crystals.
The spectra were acquired in a quasi-backscattering geometry from the ab surfaces, cleaved and transferred into the cryostat in nitrogen environment immediately prior to each cool down. 
About 10\,mW of the laser power was focused into $50\times 50\,\mu m^2$ laser spot. 
Scattered photons were collected and analyzed by a custom triple-grating spectrometer with a liquid nitrogen cooled charge-coupled device (CCD) detector. 
The secondary emission intensity, ${\rm I}_{\mu\nu}(\omega,T)$, was normalized to the laser power and corrected for the spectral response of the spectrometer and CCD, where $\mu$ ($\nu$) is the polarization of incident (collected) photon.
In a Raman process, $\rm{I}_{\mu\nu}(\omega,T)$ is related to the Raman response function, 
$\mathcal{R}_{\mu\nu}(\omega,T) = {\rm I}_{\mu\nu}(\omega,T)/[1+n(\omega,T)]$,
where $n(\omega,T)$ is the Bose-Einstein coefficient, $\omega$ is Raman shift and $T$ is temperature.

The Raman response functions for given polarizations of incident and scattered photon are defined by the rank-2 Raman tensors, which can be symmetrized according to the irreducible representations of the crystal's point group.
The scattering geometries used in this experiment are denoted as $\mu\nu=$RR, RL, XX and XY, probing $A_1+A_2$, $2E_2$, $A_1+E_2$ and $A_2+E_2$ symmetries of the $C_{6v}$ group, respectively~\cite{Cardona1982,Ovander1960}. 
$\text{R}=\text{X}+i\text{Y}$ and $\text{L}=\text{X}-i\text{Y}$ denotes the right- and left-circular polarizations, respectively, where X (Y) denotes linear polarization parallel (perpendicular) to the plane of incidence.

After subtracting luminescence contributions (Sec.\,II), the measured spectra $\rm{I}_{\mu\nu}(\omega,T)$, 
are then decomposed into $E_2$, $A_1$ and $A_2$ symmetry channels as follows:
\bea\label{eq:s1}
\mathcal{R}_{E_2}(\omega,T)&&=\frac{{\rm I}_{\rm RL}(\omega,T)}{2(1+n(\omega,T))} \nn
\mathcal{R}_{A_1}(\omega,T)&&=\frac{{\rm I}_{\rm XX}(\omega,T)-\frac{1}{2}
	{\rm I}_{\rm RL}(\omega,T)}{1+n(\omega,T)} \nn
\mathcal{R}_{A_2}(\omega,T)&&=\frac{{\rm
		I}_{\rm XY}(\omega,T)-\frac{1}{2}{\rm
		I}_{\rm RL}(\omega,T)}{1+n(\omega,T)}.
\eea

\subsubsection{Computational details}
In relation to Eq.\,(4) in the Main Text where we calculate the spin susceptibility, a momentum cutoff of $\Lambda_k=0.3$\,\AA$^{-1}$ was chosen. 
Any ambiguity that may arise due to the choice of cutoff can be subsumed into the interaction parameter $U$, thus making the physics of the appearance of the chiral spin collective modes universal. 
The threshold for the spin-flip continuum ($\omega_-$) is obtained by finding the smallest $\omega$ such that $\Pi''(\omega,T=0)\neq 0$. 
We find that $\omega_-\approx 260$ meV in the sample measured.

\subsection{II.~Photoluminescence contribution removal} 
\label{sec2}
\begin{figure}
	\includegraphics[width=7cm]{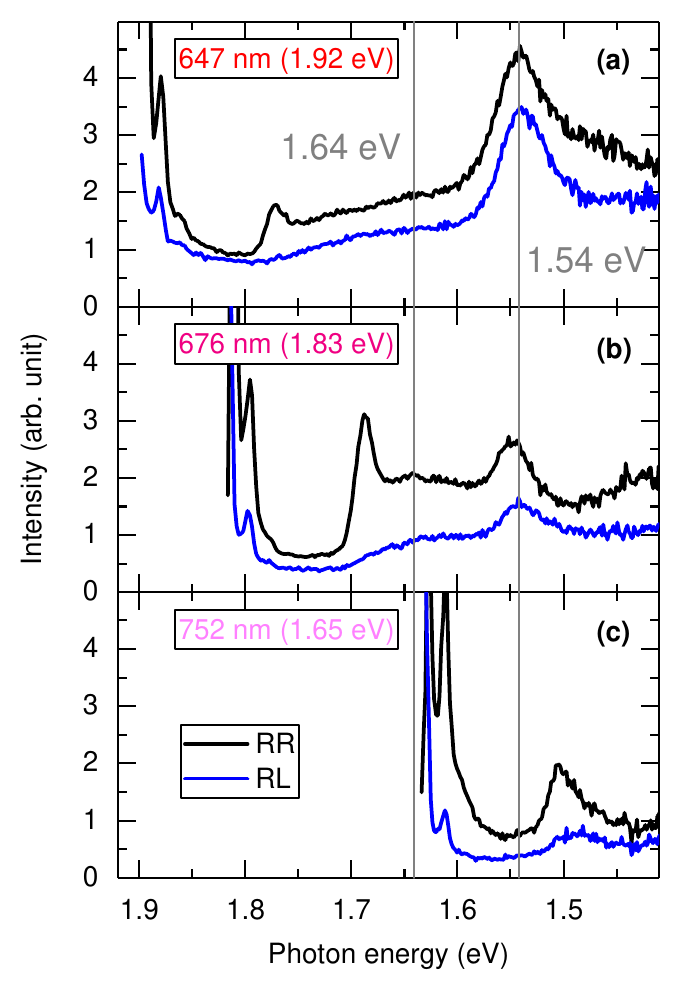}
	\caption{\label{FigS1} Raman intensity measured in RR and RL
		polarizations at 24\,K for (a)~647\,nm (1.92\,eV), (b)~676\,nm (1.83\,eV) and (c)~752\,nm (1.65\,eV) excitation
		energies, plot against scattered photon energy. The gray solid
		line marks 1.54 and 1.64\,eV, where the exciton peaks centers
		coincide for 647 and 676\,nm excitations, indicating that the
		peaks are due to photoluminescence emission rather than Raman scattering signal.}
\end{figure}

Figure~\ref{FigS1} shows the intensity of secondary emission measured for RR and
RL polarizations at 24\,K for 647, 676 and 752\,nm excitation wavelengths, plotted as function of emission photon energy. 
The exciton emission
centers at 1.54\,eV for 647 and 676\,nm excitations, and
has about the same intensity for both RR and RL scattering
geometries. 
Another weaker emission peak is observed at 1.64\,eV for both excitations.
These peaks are absent for 752\,nm excitation spectra, suggesting that the emission has a threshold of about 1.8\,eV.

To remove photoluminescence background from the measured spectra, we fit
the 1.54 and 1.64\,eV exciton peaks with a Lorentzian function, as shown by the
hatched peaks in Fig.~2 of Main Text. We also subtract a small
constant background from all spectra to account for other
photoluminescence contribution.

\subsection{III.~Transitions between surface states and bulk bands} \label{sec:supp.iii}
\begin{figure}
	\includegraphics[width=5cm]{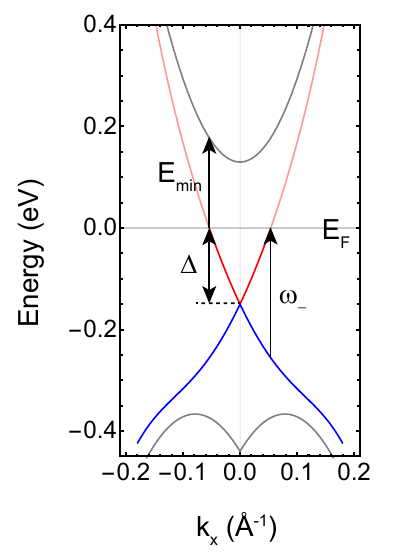}
	\caption{\label{FigS2} 
		%\textbf{Transitions
		%between surface states and bulk conduction band.}
		Band structure near the $\Gamma$ point and Fermi surface, reconstructed from ARPES measurements~\cite{Nomura2014}. 
		The blue and red lines denote the lower and upper Dirac cones, respectively, whereas the bulk bands are shown in gray. 
		In the measured sample, the Dirac point is about 150\,meV below the Fermi energy $E_F$. 
	}
\end{figure}

In relation to Fig.\,2(a) in the Main Text, we present in this section an explanation of the spectroscopic features observed for the non-resonant 521\,nm (2.38\,eV) excitation.

Figure~\ref{FigS2} 
%is a realistic illustration of 
shows the band structure of
%the
Bi$_2$Se$_3$
%band structure 
reconstructed from ARPES
measurements~\cite{Nomura2014}. 
The dispersion of the surface states are~\cite{Fu2009}:

\bea \label{eq:ss1disp}
E_{SS}(\vec{k}) =&\Delta+\frac{k^2}{2m^\ast} \pm\sqrt{v_F^2k^2+\left(\frac{2v_3}{v_F}\right)^2k^6\cos^2 3\theta} \nn
\approx& \Delta+\frac{k^2}{2m^\ast} \pm v_Fk + 2\left(\frac{v_3^2}{v_F}\right)k^5\cos^2(3\theta)~, \eea
where $\pm$ denote the upper and lower Dirac cones
and $\theta$ is the azimuth angle of momentum $\vec{k}$ with respect to the $x$ axis ($\Gamma-K$).
Fitting data in Ref.\,\cite{Nomura2014} to Eq.~\ref{eq:ss1disp} gives $m^\ast\approx 0.066\,\text{eV}^{-1}\text{\AA}^{-2}$, $v_F\approx 2.4$\,eV\AA\,, and $v_3\approx 25$\,eV\AA$^3$.
One can readily see that the 
energy of
a
direct transition  
from 
the
lower to upper Dirac cone is
$2\sqrt{v_F^2k^2+(\frac{2v_3}{v_F})^2k^6\cos^2(3\theta)}$.
In samples measured, $\Delta$ is determined by tunneling spectroscopy~\cite{Dai2016} to be about $-150$\,meV, 
therefore the Fermi momentum $k_F\approx 0.054$\,\AA$^{-1}$ along $k_x$, thus resulting
in a threshold energy $\omega_{-}\approx 260$\,meV.
%This value is substantially larger than the observed threshold energy of about 180\,meV in Fig.~\ref{Fig2}(a).

The direct transition energy between SS1 and the bulk conduction band is given by 
$\epsilon(\vec{k})=E_{CB}(\vec{k})-E_{SS}(\vec{k})$,
where $E_{SS}(\vec{k})$ is given 
by Eq.\,(\ref{eq:ss1disp}), 
and
the bulk conduction band dispersion follows a quasi-2D 
parabolic
model~\cite{Lahoud2013}:
\be \label{eq:cbdisp}
E_{CB}(\vec{k})=E_0+\frac{k_{||}^2}{2m_{||}^\ast}+\frac{k_{\perp}^2}{2m_{\perp}^\ast},
\ee
where $E_0\approx 130$\,meV is determined by $E_F$ and the relative position between SS1 and bulk conduction band minimum~\cite{Nomura2014,Dai2016}, $m_{||}^\ast\approx 0.03\,\text{eV}^{-1}\text{\AA}^{-2}$ is the in-plane 
effective mass, determined from fitting the ARPES data in Ref.~\cite{Nomura2014} 
to 
quadratic dispersion.
In the measured sample where $k_F\approx 0.054$\,\AA$^{-1}$ along $k_x$, the threshold energy $E_{\min}\approx 180$\,meV, similar to what was observed in Fig.\,2(a).

\subsection{IV.~Temperature dependence of the surface chiral spin mode} \label{sec:supp.iv}
\begin{figure}
	\includegraphics[width=7cm]{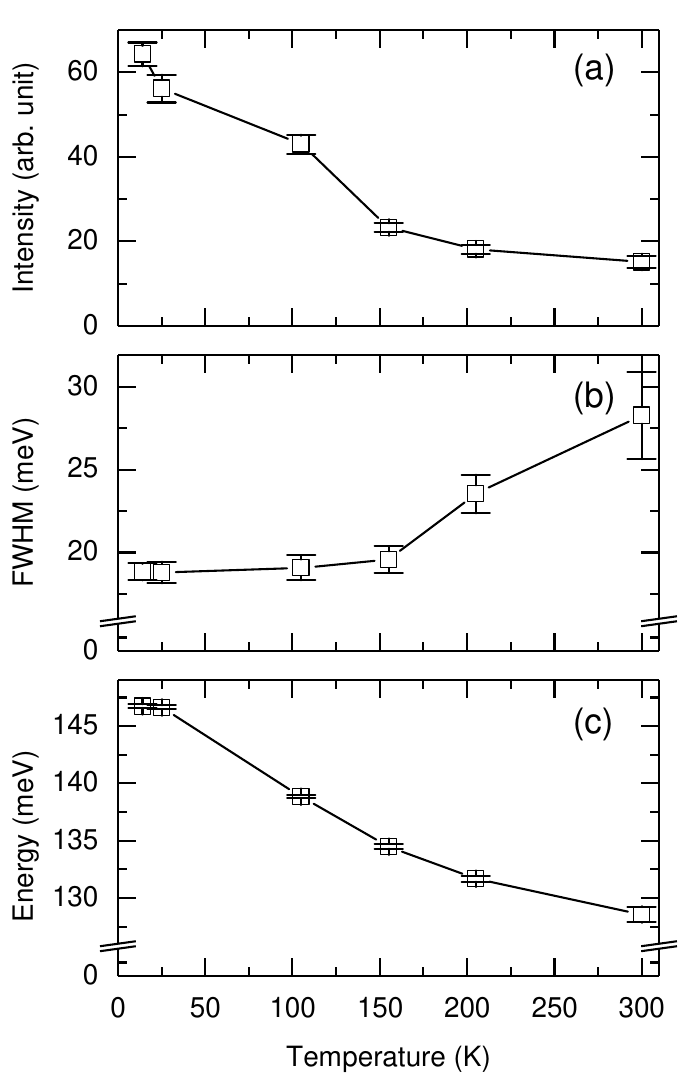}
	\caption{\label{Fig:A2gTdep} 
		%\textbf{Temperature dependence of the surface chiral spin mode.}
		Temperature dependence of (a)~the intensity, (b)~full width at half maximum (FWHM), and (c)~peak center, of the chiral spin mode [Fig.\,3(c) in the Main Text], fitted to a Lorentzian line shape.
		The error bars reflect one standard deviation of the fit.
	}
\end{figure}

Figure\,\ref{Fig:A2gTdep} 
shows
the temperature dependence of the Raman intensity [(a)], full width at half maximum [FWHM, (b)], and peak center energy [(c)]
of the chiral spin mode.
The parameters were obtained by fitting the data in Fig.\,3(c) of the Main Text to a Lorentzian lineshape.
FWHM is approximately independent of temperature for $T\leq 150$ K. This indicates that the main damping mechanism of spin waves for these temperatures is due to disorder via the D'yanokov-Perel' mechanism \cite{Dyakonov1971}. This is in line with the theoretical predictions for damping of chiral spin waves \cite{Shekhter2005,Maiti2015PRL}. At higher temperatures, inelastic scattering mechanisms, e.g., electron-electron \cite{Bir1975,Maiti2015PRL} or electron-phonon \cite{Huber1966} interactions, may also contribute to damping. However, we found that a model, which incorporates the finite-temperature effects only via thermal smearing of the Fermi functions and neglects inelastic damping mechanisms, describes the experiment rather well. The results of such model with a $T$-independent damping rate of $8$\,meV (taken as 1/2 of FWHM at $T\to 0$) are shown in Fig.~3 (d) of the Main Text. On the other hand, the fact that the measured intensity decreases with increasing temperature faster than the calculated 
%spectra
one is an indication of unaccounted spin decay channels at elevated temperatures, e.g., through interaction with surface phonons~\cite{Kung2017}.
%
%\bibliography{Kung2017}

\end{document}